\documentstyle[emulateapj]{article}

\begin{document}

\lefthead{COORAY}
\righthead{GRAVITATIONAL LENSING LIMITS ON THE AVERAGE REDSHIFT OF SUBMILLIMETER SOURCES}

\title{Gravitational Lensing Limits on the Average Redshift of Submillimeter Sources}
\author{Asantha R. Cooray}
\affil{Department of Astronomy and Astrophysics,
University of Chicago, Chicago IL 60637\\ 
E-mail: asante@hyde.uchicago.edu}
\begin{center}
{\it Received: 1998 November 19; accepted: 1998 December 16}
\end{center}


\begin{abstract} 
The submillimeter universe has now been explored with the Submillimeter
Common User Bolometer Array (SCUBA) camera 
on the James Clerk Maxwell Telescope, and a claim
has been made to the presence of a new population of optically unidentified
starforming galaxies at high redshifts $(z \gtrsim 3)$. 
Such a population dramatically alters current views on the star
formation history of the universe as well as galaxy formation and
evolution. Recently, new radio identifications of the
Hubble Deep Field submm sources have led to the suggestion
that some of these sources are at low redshifts, however, 
submm source redshift distribution is still not well determined.  
Here, we present an upper limit
to the average  redshift $\langle z \rangle$ by comparing the
 expected number of gravitationally lensed 
submm sources due to foreground cluster potentials to current observed 
statistics of such lensed sources. The upper limit depends on the
cosmological parameters, and at the 68\% confidence level,
$\langle z \rangle <$ 3.1, 4.8, 5.2, or 8.0 for ($\Omega_m,\Omega_\Lambda$)
values of (0.3,0.7), (0.5,0.5), (0.3,0.0) or (1.0,0.0) respectively.
These upper limits are consistent with redshift distribution 
for 850 $\mu$m sources implied by starformation history models 
based on measured background radiation at far-infrared and submm wavelengths.

\end{abstract}



\section{Introduction}

Several deep surveys have now been carried out at a wavelength of 850 $\mu$m
using the Submillimeter Common User Bolometer Array (SCUBA) camera 
on the James Clerk Maxwell Telescope. These surveys have led to
the identification of a sample of submm sources with
flux densities in the range 2 to 10 mJy at 850 $\mu$m
with a surface density of $\sim$ 1000 deg$^{-2}$.
Given that the SCUBA resolution is relatively poor and that
there are roughly 20 optical galaxies down to
R band magnitude of 26 out to a redshift of 4 in the SCUBA beam,
definite optical identification of detected submm sources has not
been possible. 

Under the assumption that the submm source population is
at redshifts greater than $\sim$ 1, Hughes et al. (1998) identified
submm sources in the Hubble Deep Field (HDF; Williams et al.
1996) with galaxies in the redshift range of 2.5 to 4. 
The implied starformation rate in this redshift range based on the
submm flux densities was order of an magnitude greater than what
was previously calculated using optical, ultraviolet and infrared data. 
Recently, submm counterparts in the HDF were
questioned by Richards (1998) based on new 1.4 GHz radio counterparts to
these submm sources. These radio identifications suggest that 
the optical counterparts
to most of the HDF submm sources are at redshifts lower than
what were previously suggested by Hughes et al. (1998). Such
a low redshift distribution dramatically changes
the previous claim for a substantially higher starformation rate at 
redshifts greater than $\sim$ 2. 

Based on spectroscopic redshifts of submm sources detected
by Eales et al. (1998) in the fields containing previous 
Canada-France-Redshift Survey (CFRS),
Lilly et al. (1998) concluded that most of the submm
sources are at redshifts less than $\sim$ 1. 
Such a low redshift distribution 
is also compatible with optical counterparts 
based on archival Hubble Space Telescope observations for 
submm sources detected towards galaxy clusters (Smail et al. 1998a; 
hereafter S98). We refer the reader to Smail et al. (1998b) for a recent 
review on published deep SCUBA surveys at 850 $\mu$m. These surveys include
the HDF (Hughes et al. 1998), CFRS fields (Eales et al. 1998),
the Lockman hole and Hawaii Survey Fields (Barger et al. 1998),
and lensing clusters (Smail et al. 1997; S98).

Based on a combined analysis of the submm and far-infrared source counts
combined with measured values for the background intensities at millimeter,
submm, and far-infrared wavelengths, Blain et al. (1998) suggested
that 90\% of the 850 $\mu$m sources are at redshifts below 3.8 and 8.2.
The same analysis also concluded that there is no peak in the
starformation rate between redshifts of 1 and 2, as previously
suggested by the optical and ultraviolet data, but rather
the starformation increases as $(1+z)^4$ till a redshift of
$\sim 1$, and remains constant thereafter. If most of
the submm sources are at high redshifts ($z \gtrsim 4$), then
deep  submm wavelength surveys  should begin to recover instances 
of gravitationally lensed submm sources, 
as the probability for lensing increases with redshift.
If the redshift distribution of background and foreground
sources are known, the number of lensed sources in a given survey 
can be used to constrain cosmological
world models, especially the cosmological constant (e.g., Turner et al. 1990).
If the background source redshift distribution is unknown,
lensed source statistics can be used to study background source redshifts
using prior knowledge on the cosmological world model and foreground lenses. 

In Cooray (1999a; hereafter C99), we studied the former case and
suggested the possibility of constraining cosmological
parameters based on observed statistics of lensed sources at
850 $\mu$m towards a sample of galaxy clusters. 
This approach is quite similar to the one taken in literature to 
constrain the present day value of the cosmological constant based
on lensed source statistics, such as quasars (e.g., Kochanek 1996; Chiba \&
Yoshi 1997), radio sources (e.g., Falco et al. 1998; Cooray 1999b) and 
 luminous optical arcs towards clusters (e.g.,
Bartelmann et al. 1998; C99). Blain (1997) studied
gravitational lensing at submm wavelengths, including
statistics of lensed submm sources due to foreground galaxies.

Here, we consider the possibility of constraining the background
source redshift distribution  based on known properties of a sample
of submm sources gravitationally lensed by cluster potentials (S98)
and an assumed cosmological model. To obtain information on the unlensed
sources down to the same flux level, we use the source counts
from Barger et al. (1998), 
Eales et al. (1998), Holland et al. (1998),  Hughes et al. (1998) and
Smail et al. (1997, 1998b).
In \S~2 we discuss our calculation and its inputs.
in \S~3 we present our resulting constraints on the redshift
distribution of submm sources and discuss
our constraints in the context of current 
studies on the submm sources and their contribution to the
starformation history of the universe.
We follow the conventions that the Hubble
constant, $H_0$, is 100\,$h$\ km~s$^{-1}$~Mpc$^{-1}$, the present mean
density in the universe in units of the closure density is $\Omega_m$,
and the present normalized cosmological constant is
 $\Omega_\Lambda$. In a flat universe, $\Omega_m+\Omega_\Lambda=1$.

\section{Gravitational lensing calculation}

Our calculation follows C99 in which we calculated the expected number of
luminous optical arcs, radio sources and submm galaxies towards
galaxy clusters as a function of cosmology.
Here, we prescribe the foreground cluster population to be similar to
what was observed by S98 and model them as singular
isothermal spheres (SIS) with  velocity 
dispersion $\sigma$. 
In general, SIS models underestimate the
number of lensed sources, when compared to complex cluster potentials with
substructure (e.g.,  B\'ezecourt 1998). This leads to a systematically lower
number of lensed sources than expected from true complex potentials
and a higher upper limit on the
redshift distribution of background sources. 

In order to evaluate distances, we use the analytical filled-beam approximation
(see, e.g., Fukugita et al. 1992) and calculate the probability, 
$p(z,\Omega_m,\Omega_\Lambda)$,  for a source
at redshift of $z$ to be strongly lensed given a set of cosmological
parameters $\Omega_m$ and $\Omega_\Lambda$. Following C99 (see,
also, Cooray, Quashnock \& Miller 1999 and Holz, Miller \& Quashnock 1999)
the number of expected lensed sources, $\bar N$, 
towards the survey volume containing foreground lensing clusters is:
\begin{eqnarray}
\bar N = & \int F(z_l) p(z,\Omega_m,\Omega_\Lambda) B(f,z) dC(z) \nonumber \\
 & \equiv \int \tau(z, \Omega_m,\Omega_\Lambda) dC(z) \;            
\end{eqnarray}
where $C(z)$ is the redshift distribution of submm sources such that
$C(z)$ is the fraction of sources with redshifts less than $z$,
$B(f,z)$ is the magnification bias for submm sources at redshift $z$
with observed flux density at 850 $\mu$m of $f$ (see,
Kochanek 1991), and
$F(z_l)$ is the effectiveness of clusters at redshifts
$z_l$ in producing lensed sources. This nondimensional
parameter can be written as (Turner, Ostriker\& Gott 1984):
\begin{equation}
F(z_l) = 16 \pi^3 n(z_l) \left( \frac{\sigma}{c}\right)^4 R_0^3,
\end{equation}
Here, $R_0=c/H_0$, and $n(z_l)$ is the number density of clusters
with the velocity dispersion $\sigma$ at redshift $z_l$. 

In general, 
detailed knowledge either on the luminosity function or the
flux distribution is required to calculate 
the magnification bias.
However, both these quantities are currently not  known for
the submm source sample. Instead of individual magnification biases, we 
use current estimates on the submm sources number counts to obtain
an average value.
If the number counts of unlensed sources, $n_{ul}$,
 with flux densities greater than $S_\nu$ 
towards a given area can be written as $n_{ul} \propto S_\nu^\alpha$, then
magnification due to gravitational lensing by an amplification $A$ modifies the
counts  as:
\begin{equation}
n_{l} \propto A^{-(1+\alpha)} S_\nu^\alpha,
\end{equation}
where $n_l$ is the lensed source counts.
 The average magnification
bias is simply the ratio of lensed to unlensed counts down to a flux density
$S_\nu$:
\begin{equation}
\langle B \rangle =  \frac{n_l}{n_{ul}} = A^{-(1+\alpha)}. 
\end{equation}
Under the SIS scenario, 
the probability distribution for amplifications is $P(A) = 2/(A-1)^3$,
and the minimum amplifications is $A_{\rm min}=2$.
The average amplification for a sample of lensed sources is 3. 
This average value is 
consistent with the distribution of amplifications for the submm sources
 based on detailed modeling of individual cluster potentials: 
$1.3^{+5.0}_{-0.5}$ (S98). Since none of the observed lensed
sources are heavily amplified due to foreground potentials and that the 
amplification distribution is compatible with the SIS average,
our use of SIS model to describe foreground clusters should not
affect the results greatly. 

Following Smail et al. (1998b), we parameterized submm source counts 
at 850 $\mu$m as:
\begin{equation}
N(>S) = (7.7 \pm 0.9) \times 10^3 \left(\frac{S}{1\, {\rm mJy}}\right)^{-(1.1 \pm 0.2)},
\end{equation}
where the uncertainties are the 1$\sigma$ errors. 
The slope $\alpha$ is $-(1.1 \pm 0.2)$, and thus, 
the average magnification bias, $\langle B(f,z) \rangle$, 
ranges from 0.9 to 1.4.
This estimate for the magnification bias for 850 $\mu$m sources with flux
densities in the range of 0.5 to 10 mJy is slightly lower 
than what was previously considered (e.g., Blain 1997).  
For the purpose of this calculation, where we are
only interested in an upper limit to the redshift distribution,
we apply the lowest possible amplification bias to all background sources.
This leads to an underestimated lensing rate
and an  overestimated upper limit on the background source redshift 
distribution.  

Since the redshift distribution of submm sources, $C(z)$, is unknown,
we calculate the observed number of lensed sources as a function of
$\langle z \rangle$, the effective average redshift under the assumption
that all sources are at this redshift:
\begin{equation}
\bar N = \langle F(z_l) \rangle p(\langle z \rangle,\Omega_m,\Omega_\Lambda) \langle B(f,z) \rangle. 
\end{equation}
 This approach is essentially
similar to the one taken by Holz et al. (1999) to calculate an upper limit
to the redshift distribution of gamma ray bursts based on observed
lensing statistics. The assumption of an average redshift is utilized to 
parameterize our ignorance of the  submm source redshift distribution 
(see, Holz et al. 1998 for a discussion).

The S98 sample contains observations of 7 clusters with an effective total area
surveyed of 0.01 deg$^2$. This sample lies in redshifts between 0.19 
and 0.4. The highest probability for foreground clusters to lens
background sources occurs in the redshift range of 0.2 to 0.7, with
some slight dependence on the cosmological parameters (see, e.g.
Fig.~2 in Bartelmann et al. 1998).
For the purpose of this calculation, we calculate an average
$F$ parameter, $\langle F(z_l) \rangle$, based on the number density
of clusters in this redshift range. Since the S98 cluster sample contains
some of the well known massive clusters, we use a lower limit on the mass
distribution determined by the observed velocity dispersions of these
seven clusters, and assuming
virial theorem for galaxy clusters (see, C99 for further details). 
The distribution of velocity dispersion for the 7 clusters 
observed by S98 has a mean of $1150 \pm 310$ km s$^{-1}$.
The number density of clusters 
was calculated based on the Press-Schechter (PS;
Press \& Schechter 1974) theory with normalization based on the
local cluster temperature function. Our PS calculation follows
Bahcall \& Fan (1998), with normalizations for $\sigma_8$ presented
therein. We calculated number density of clusters with virial masses above 
$7.5 \times 10^{14}\, M_{\sun}$,
corresponding to the observed velocity dispersion based on virial theorem,
as a function of redshift and $\Omega_m$. 
In order to account for the
uncertainty in $\langle F(z_l) \rangle$ due to measurement errors and 
our assumption of an average value,
we allow for an overall uncertainty 
of 40\%.

\section{Constraints on the redshift distribution}

Using observational results from S98, we assume that number 
of lensed sources towards seven
clusters is 10 down to a flux density limit of $\sim$ 6 mJy. 
Thus, the lensing rate per cluster down to
6 mJy is about 1.42, which is roughly 
a factor of 2 higher than the lensing rate for luminous optical arcs with
amplifications greater than 10 down to a I band magnitude of 22 
(C99; see, also Le F\`evre et al. 1994).
Some of the currently presumed lensed submm sources are likely to be cluster
member or foreground sources. Several such sources have already been detected
within the current S98 sample which have been
identified with cluster cD galaxies (Edge et al. 1998) and foreground
spiral galaxies at low-redshifts (S98).
However, for the purpose of this paper, where we are only interested
in a statistical upper limit to average redshift, we can safely take 10 as the
observed number of lensed sources. As before, the only effect of such
an assumption is to
increase the derived upper limit from the true value.

We also ignore biases in the S98 cluster sample and assume
it as a random and a fair sample of clusters on the sky. The clusters imaged
by S98 are some of the well known massive clusters, towards which
the optical lensing rate is somewhat higher than the average value.
This is primarily due to the fact that some of these clusters are found with
substructures and bimodal mass distributions, producing enhanced
potentials for gravitational lensing. Here again, the systematic bias
is such that the  observed number of sources 
is an overestimate of the average number, and the derived
 upper limit on the background redshift
is an overestimate from the true upper limit. 

We calculated the expected number of lensed sources down to
a flux density of 6 mJy  at 850 $\mu$m as a function of redshift
for different cosmological models. We vary the redshift of the background  
sources, assuming that all of them are at the same redshift, and
calculate the expected number of lensed sources towards
clusters on the whole sky. Based on cluster abundance from PS theory,
we convert the number of lensed sources to an average
value expected towards seven clusters, such as to mimic S98
observations. We then compare this number to the observed number,
and following Cooray et al. (1999) we consider
a Bayesian approach to calculate an upper limit to the average
redshift of submm sources, under the assumption of a cosmological model.
The likelihood ${\cal L}$ --- a function of $\langle z \rangle$, $\Omega_m$
and $\Omega_\Lambda$ ---  is the probability of the data, 
given $\langle z \rangle$, $\Omega_m$ and 
$\Omega_\Lambda$. The likelihood for $n$ observed sources 
 when $\bar N$ is expected towards seven clusters is given by:
\begin{eqnarray}
\langle {\cal L}(n)\rangle = & \prod_{j=0}^n \tau(\langle z \rangle) 
\times e^{-\bar N(\langle z \rangle)} \nonumber \\
& \times 
\left(1 + \sigma_{\cal F}^2 \left[ \frac{\bar N(\langle z \rangle)^2}{2}-n\bar N(\langle z \rangle) +\frac{n(n-1)}{2}\right]\right) \; .
\end{eqnarray}
Here, $n$ is the observed number while $\bar N(\langle z \rangle)$ is the
expected number of lensed sources 
when background sources are at $\langle z \rangle$
when $\Omega_m$ and $\Omega_\Lambda$ is given.
We have also taken into account the uncertainty in $\langle F(z_l) \rangle$ 
by defining  $\sigma_{\cal F}$ to be 0.4, allowing for a 
40\% uncertainty.
This factor is an overall correction to the
expected lensing rate, due to a systematic uncertainty in $\langle F(z_l) 
\rangle$.

In Table~1, we list the derived 68\% and 95\% upper limits on the redshift
distribution of submm sources to produce the current observed
statistics. If $\Omega_m=0.3$ and
$\Omega_\Lambda=0.7$, as suggested by various cosmological probes,
including a combined analysis of the Type Ia supernovae at
high redshifts (e.g., Riess et al. 1998) and cosmic microwave
background anisotropy data (e.g., White 1998; Lineweaver 1998),
galaxy cluster evolution (e.g., Bahcall \& Fan 1998) and baryonic
fraction in galaxy clusters (e.g., Evrard 1997), then
the 68\% upper limit on the redshift distribution is 3.1.
If $\Omega_\Lambda=0.0$ and the universe is open with
$\Omega_m$ of 0.3, then this upper limit increases to 5.2.

Our upper limits on the average redshift can be directly compared to redshift
distribution derived by Blain et al. (1998) based on the modeled
starformation history using submm and far-infrared number counts and
background radiation measurements. The authors derived that
90\% of the sources at 850 $\mu$m will lie below the redshift range of 3.8
to 8.2, with median redshift in the range of 2.4 to 4.4. Our average
upper limit $\langle z \rangle$ of 3.1 for $\Omega_m=0.3$ and
$\Omega_\Lambda=0.7$ is in agreement with such a distribution.
Blain et al. (1998) calculation on the redshift distribution
assumes a cosmology of $\Omega_m=1.0$ and $\Omega_\Lambda=0.0$.
This is the same cosmology for which we have the weakest upper limit
with  $\langle z \rangle$ of 8.0 at the 68\% confidence. 
The derived redshift limit here also agrees with suggested redshift ranges
by S98 using individual colors of plausible identifications.

\section{Uncertainties and Systematic Effects}

We have assumed that the S98 sample is a fair sample of galaxy clusters,
and have considered it to obtain an average number of observed
lensed sources due to foreground galaxy clusters. This assumption is likely
to be false given biases and systematic effects in the cluster 
sample selection.  Our treatment of pointed cluster observations as a series 
of random untargeted observations is likely to create an additional 
systematic bias, but such a bias is not expected to
underestimate the current upper limit.
We have also considered a low value for the magnification bias such
that the upper limit on background source
redshift is overestimated. If, for example, the true magnification
bias is 1.4, then the upper limit on $\langle z \rangle$
decreases to 2.6 from 3.1 in a cosmology of $\Omega_m=0.3$ and
$\Omega_\Lambda=0.7$. 

Other uncertainties include the determination of
cluster abundances given systematic and statistical
uncertainties involved with the PS calculation, resulting from errors
due to $\sigma_8$ etc. We have tried to compensate for such errors
by considering a 40\% statistical uncertainty in the derivation of
$\langle F(z_l) \rangle$. In general, it is likely that we
have overestimated the upper limit, since most of the
systematic effects tend to bias our results such that we
 underestimate the expected lensing rate.

\section{Summary and conclusions}

We have derived upper limits on the redshift distribution of submm sources by
comparing statistics of lensed sources towards a sample
of galaxy clusters to unlensed sources. Our derived
limits depends on cosmology, and if $\Omega_m=0.3$ and $\Omega_\Lambda=0.7$,
as currently suggested by various cosmological probes, at the 68\% level
the average redshift of submm sources is less than 3.1.
Such an upper limit is consistent with the redshift
distribution predicted for submm sources based on starformation models,
where starformation history remains constant beyond a redshift of
1.5, using observed far-infrared and submm background radiations. 
The derived upper limit on the average redshift is also consistent with
suggested redshift ranges based on colors of plausible optical identifications
for submm sources detected towards cluster potentials.

\acknowledgments

I would  like to thank the anonymous
referee for  constructive comments on the paper.

\begin{deluxetable}{rrcc}
\tablewidth{30pc}
\tablenum{1}
\tablecaption{Upper Limit on the average redshift $\langle z \rangle$ for
different cosmologies.}
\tablehead{
\colhead{$\Omega_m$} &
\colhead{$\Omega_\Lambda$}    &
\colhead{ $\langle z \rangle$ upper limit 68\% } &
\colhead{ $\langle z \rangle$ upper limit 95\% }}
\startdata
0.3 & 0.7 & 3.1 & 4.9 \nl
0.5 & 0.5 & 4.8 & 7.7 \nl
0.3 & 0.0 & 5.2 & 8.1 \nl
1.0 & 0.0 & 8.0 & 12.2 \nl 
\enddata
\end{deluxetable}

\end{document}